\documentclass[epj]{svjour}
\usepackage{graphics}
\usepackage{amsmath}
\usepackage{graphicx,epsfig,psfrag}
\usepackage{amssymb}

\newcommand{\OLD}[1]{{\bf OLD RM}}

\begin{document}
\title{Wilson chains are not thermal reservoirs}
\author{Achim Rosch
}                     
%
%
\institute{Institute for Theoretical Physics, University of Cologne, 50937
Cologne, Germany}
\date{Received: date / Revised version: date}
%
\abstract{ Wilson chains, based on a logarithmic discretization of a
continuous spectrum, are widely used to model an electronic (or bosonic)
  bath for Kondo spins and other quantum impurities within the
  numerical renormalization group method and other numerical
  approaches. In this short note we point out that Wilson chains can
  {\em not} serve as thermal reservoirs as their 
  temperature changes by a number of order $\Delta E$ when a finite
  amount of energy $\Delta E$ is added. This proves that for a large class of
  non-equilibrium problems they cannot be used to predict the
  long-time behavior.
\PACS{
      {71.10.-w}{Theories and models of many-electron systems}   
     } 
} 
\maketitle

In 1975 Wilson pioneered \cite{wilson} the ``numerical renormalization group'' (NRG)
approach to solve the Kondo model. This approach is able to provide
numerically exact solutions to the Kondo problem, probably the most
important and most influential quantum impurity model.
The NRG is by now widely used for a large
range of fermionic and bosonic quantum impurity problems both at zero and finite
temperature \cite{review}.

A quantum impurity problem is defined by some  local degree
of freedom coupled to a (most often non-interacting) bath. In the case
of the classical Kondo model this bath is given by an infinite system
of electrons with a constant density of states at the Fermi energy.

Wilson's numerical solution consists of two main steps. First the bath
is discretized using energies defined on a logarithmic grid.
\begin{eqnarray}
\epsilon_{\pm |n|}=\pm \frac{D}{\Lambda^{|n|}}
\end{eqnarray}
This logarithmic grid is optimized for the logarithmic
renormalization group flow characteristic for the Kondo model and
allows to capture even the exponentially small energy scales
characteristic for the Kondo problem. In
praxis \cite{review} it is implemented via the so-called Wilson chain,
a chain of local sites coupled by nearest neighbor hopping rates which
decay exponentially in the distance from the quantum
impurity.
The second step, the actual renormalization group transformation,
strongly builds on this logarithmic discretization \cite{review} but it
plays no role for the following discussion.

An exciting recent development is that it is now possible to study
quantum impurity problems numerically in non-equilibrium using
NRG \cite{theo,AndersTDNRG,anders08,hofstetterTDDMRG,anders11}.
Especially F. B.  Anders and A. Schiller pioneered
powerful methods to study quantum quenches \cite{AndersTDNRG} and to
investigate steady-state transport using many-particle scattering
states \cite{anders08}.
Also other numerical methods, most notably time-dependent density
matrix renormalization group have been 
applied to Wilson chains \cite{DMRGwithWilson}.

NRG and the use of Wilson chains has been extremely
successful \cite{review}. Therefore it is a relevant question to
understand whether there are limitations to this approach. In this
short note we briefly discuss a very severe limitation: Wilson chains
can not be used as thermal reservoirs. We argue that this implies that
for certain classes of non-equilibrium problems Wilson chains give
wrong results.

A reservoir is by definition an infinite system which can be used to
fix the chemical potential and the temperature. More precisely, when
adding a finite amount of particles or a finite amount of energy, the
chemical potential and the temperature do not change. While in
experimental systems reservoirs are never really infinite, in praxis
the large number of electrons, for example in a metallic contact
attached to a quantum dot, provides an almost ideal realization of a
reservoir.  

Within this definition, the Wilson chain can serve as a reservoir for
particles as one can check by calculating $\frac{d N}{d \mu}$, for the
non-interacting Wilson chain in the absence of the quantum impurity
\begin{eqnarray}
\left. \frac{d N}{d \mu}\right|_{\mu=0}&=&- 2 \sum_{|n|<N_{\rm sites}}  f'(\epsilon_n) \approx 
\sum_{|\epsilon_n|<T,
|n|<N_{\rm sites}}  \frac{1}{2 T} \nonumber \\
&\approx&
\frac{N_{\rm sites}}{T}
 \to \infty.
\end{eqnarray}
Here $N$ is the total number of particles,
$f(\epsilon)=1/(e^{(\epsilon-\mu)/T}+1)$ is the Fermi function and the
factor $2$ accounts for the spin degeneracy. $\frac{d N}{d \mu}$
diverges with the number of sites in the Wilson chain implying that in
practical calculations the chemical potential stays constant when, for
example, $N$ increases by 1.

In contrast, the total energy of the Wilson chain and its derivative with
respect to temperature $T$ remain finite in the limit $N_{\rm
  sites}\to \infty$ for finite discretization parameter $\Lambda>1$.
\begin{align}
\frac{d E}{d T}&=&- 2 \sum_{n}  \frac{\epsilon_n^2}{T} f'(\epsilon_n)
\approx 4 \int_{-\infty}^{\infty} dn \frac{ \Lambda^{-2 n} e^{\Lambda^{-n}}}{(1+e^{\Lambda^{-n}})^2} =\frac{4 \ln
  2}{\ln \Lambda} \label{E}
\end{align}
In practical calculations $\Lambda$ is never very close to $1$ (an
often used value is $\Lambda=2$) therefore $dE/dT$ is a number of
order 1!

This has severe practical implications for non-equi\-li\-brium
situations. Consider, for example, a quantum quench where at time
$t=0$ suddenly a hybridization of strength $\Gamma$ is switched on,
which couples an Anderson impurity to a Wilson chain at $T=0$. The
following time evolution conserves energy. The ground state energy of
the system with finite hybridization is approximately
lower  by $\Gamma$ than the energy of the initial state. Assuming that the system
equilibrates, Eq.~(\ref{E}) implies that final state has a huge
temperature of order $\Gamma \ln \Lambda$, much larger than the Kondo
temperature. In contrast, an Anderson impurity coupled to a real
reservoir, where $dE/dT$ diverges with system size, relaxes to the
$T=0$ groundstate in the thermodynamic limit.  While in real systems
the energy can be transported away from the quantum impurity to
infinity this is not possible for the Wilson chain due to its finite
heat capacity. Our argument proves that in the long time limit a wrong
result is obtained for all quantum quenches where the energy of the
initial configuration is higher than the ground state
energy by an amount of the order of the energy scale one is interested
in.  With Eqn. (\ref{E}) one can estimate quantitatively the
heating of the quantum impurity.  This can serve as a quantitative
estimate of the backreaction to the quantum impurity which arises as
the Wilson chain is not able to transport energy away to infinity.

It is less clear to what extent the short-time dynamics
is affected but even in this case it would not be surprising if the
problem to transport energy away is of relevance even shortly after a
quantum quench. Recently, P. Schmitteckert \cite{schmitteckert} made a
closely related observation by studying the time-evolution of a wave
packed on a Wilson chain: the wave packed piles up in a ``Wilson
Tsunami'' and is not able to reach the end of the chain.

It is an interesting question whether and to what extent the finite
heat capacity of Wilson chains is also of importance when using
scattering states to treat steady-state non-equilibrium
\cite{anders08}. Consider, for example, a quantum dot in the Kondo
regime in the presence of a finite bias voltage \cite{rosch}. In this
case the finite product of voltage and current implies that power is
pumped into the system. This power is ultimately dissipated in
infinite reservoirs where the precise form of the dissipation
mechanism is usually not important as the relaxation occurs far way
form the impurity. In the case of an infinite Wilson chain where
energy cannot be added without modifying the bath, one can suspect
that (as in the case of a quantum quench) the interacting
non-equilibrium situation cannot be treated properly. Interestingly,
present NRG implementations seem not to be able to recover the result
\cite{splitting,rosch} that the Kondo peak in the spectral function
splits in the presence of a bias voltage which exceeds the Kondo
temperature. This result is well established by (renormalized)
perturbation theory and controlled one-loop renormalization group
\cite{rosch} calculations. Possibly, the numerical results which have
been obtained for the Anderson model where not sufficiently deep in
the Kondo regime to see the splitting. Alternatively, it is also
possible that the use of Wilson chains is the origin of the
problem. Note, however, that also other numerical
approaches which are not relying on Wilson chains have presently
problems to obtain the splitting \cite{komnik}.

In contrast to the fermionic Wilson chain, for bosons the 
heat capacity of the Wilson chain is infinitely large due to the
possiblity to put a large number of bosons into the states with very
low energy (replacing the plus by a minus sign in the integral of
Eq.~(\ref{E}) leads to a divergence).
In practical NRG implementations \cite{review}, however,
only  a finite number of bosons is taken into account which renders
$dE/dT$ again finite.

In conclusion, it seems that the use of Wilson chains for non-equilibrium problems is
a much more dangerous approximation compared to the equilibrium
case as Wilson chains cannot take up finite amounts of energy without
changing their properties.
 
\acknowledgement I thank F. B. Anders, R. Bulla, Ch. Sch\"utte,
E. Sela, K. Rodriguez for useful discussions. This work
was supported by SFB 608 of the DFG.

\end{document}